\documentstyle[aps,prl,psfig]{revtex}

\newcommand{\be}{\begin{equation}}
\newcommand{\ee}{\end{equation}}

\newcommand{\bex}{\begin{eqnarray}}
\newcommand{\eex}{\end{eqnarray}}
\begin{document}

\title{Observables in Relativistic Quantum Mechanics}
\author{R. Srikanth\thanks{e-mail: srik@iiap.ernet.in}}
\address{Indian Institute of Astrophysics, \\
Koramangala, Bangalore- 34, Karnataka, 
India.}
\maketitle
\date{}


\begin{abstract} 
We propose a quantum clock synchronization protocol in which Bob makes a
remote measurement on
Alice's quantum clock via a third qubit acting as its proxy. It is shown
that the resulting correlations are dependent on the choice of the hypersurface
along which Bob's measurement of the proxy is deemed to collapse the entangled
state vector. A proper
characterization of observables in relativistic quantum mechanics is
therefore constrained by relativistic covariance as well as causality.
\end{abstract}

\parskip 3pt

~~~~~~~~

Whereas the unitary evolution of states in quantum mechanics (QM) is given
by covariant equations of motion \cite{bjo64}, the collapse of the
state vector is manifestly not.
Traditionally, the observable is a Hermitian operator defined on an
equal-time hypersurface in the inertial reference frame of the observer.
This hypersurface corresponds to the simultaneous
state-vector collapse of the observed state. Although this does
not necessarily violate causality, it is problematic from the viewpoint of 
special relativity since simultaneity is not Lorentz invariant 
\cite{bec01}. Covariant formalisms to describe
state vector reduction have been discussed in Refs. \cite{hk}.
The problem of how operations in QM are constrained
by causality has been considered by a number of authors
\cite{aa}. Issues pertaining to the causality
and localizability of superoperators on bipartite systems have been dealt with
by Beckman et al. \cite{bec01} and Eggeling et al. \cite{egg01}. 

The process of narrowing down of the probability distribution of the measured
observable that accompanies a measurement is called reduction \cite{bra94}.
Whether reduction reflects only a change in our knowledge of the system, or
an objective alteration of the system such as in the manner described in Refs. 
\cite{om1,om2,zeh_joo_zur}, or an abrupt nonlocal collapse of the wavefunction
precipitated by a classical observer,
are important and difficult questions of physical 
and interpretational interest that lie beyond the scope of the present work.
We are concerned only with the effective picture of how the quantum 
description of a system should change in response to measurements in a 
special relativistic setting. 


To this end, we consider two (for simplicity:) stationary 
observers, Alice (located at $x = a$) and Bob (at $x = b$),
sharing pairs of entangled qubits.
Alice's qubit (``$A$") is a quantum clock, i.e., it has nondegenerate 
eigenstates, governed by the Hamiltonian
$H = \omega (|0\rangle\langle 0| - |1\rangle\langle 1|)$
with energy eigenvalues $\pm \omega$ (setting $\hbar = 1$). 
The set-up is similar to the quantum
clock synchronization protocol \cite{josz}, 
except (for one) that Bob's qubit (``$B$") states remain degenerate. 

At time $t = 0$, Alice and Bob share the singlet state
\begin{equation}
|\Psi(0)\rangle_{AB} = 
\frac{1}{\sqrt{2}}(|01\rangle - |10\rangle).
\end{equation}
After lapse of time $t$ (which is also proper time in Alice's rest-frame), the
above state has evolved to:
\begin{eqnarray}
\label{diftau}
|\Psi(t)\rangle_{AB}  &=&   
\frac{1}{\sqrt{2}}(e^{i\omega t}|01\rangle 
- e^{-i\omega t} |10\rangle) \nonumber \\
&=& |{\small +}\rangle|\psi^+(t)\rangle ~-~ 
|{\small -}\rangle|\psi^-(t)\rangle ,
\end{eqnarray}
where $|\pm\rangle \equiv (1/\sqrt{2})(|0\rangle \pm |1\rangle )$ and
$|\psi^\pm(t)\rangle \equiv \pm i\sin\omega t|\pm\rangle \mp 
\cos\omega t|\mp\rangle$. Suppose 
Alice measures in the $+/-$ basis at time $t_a$. Thereby she prepares Bob's
qubit in one of the states $|\psi^{\pm}\rangle$
instantaneously as seen from her frame. Subsequently, Bob measures qubit $B$
in the $+/-$ basis. Alice classically communicates her result to Bob.
Working on a large number of pairs of qubits, Bob expects to verify the 
joint $CB$ (and hence $AB$) measurement probabilities
\begin{eqnarray}
\label{probs}
P(+,+) &=& P(-,-) = \frac{1}{2}\sin^2\omega t_a ,\nonumber \\
P(+,-) &=& P(-,+) = \frac{1}{2}\cos^2\omega t_a.
\end{eqnarray}
We note that the correlations in Eq. (\ref{probs}) depend only on time 
$t_a$ of Alice's measurement and
are independent of the choice of the collapse hypersurface.
Therefore, as such QCS is unsuitable to test for possible dependence of
correlations on the reduction hypersurface.

Alternatively, Bob can make a remote measurement on
Alice's qubit via a proxy. By prior arrangement, Alice uses a third particle, 
a degenerate qubit denoted $C$. This qubit is made to locally interact 
with $A$, the outcome of which is represented by the 
unitary operation $U$ given by:
\begin{equation}
(\alpha|+\rangle + \beta|-\rangle)_C\otimes|\pm\rangle_A  
~ \stackrel{U}{\longrightarrow} ~ 
(\alpha|\pm\rangle + \beta|\mp\rangle)_C\otimes|\pm\rangle_A ,
\end{equation}
where $|\alpha|^2 + |\beta|^2 = 1$. Per the protocol, 
Alice begins with $C$ in the 
state $|+\rangle$. After the interaction, she
classically sends Bob the proxy qubit $C$.
The state of the 3-qubit system now is:
\begin{equation}
\label{huhu}
|\Upsilon\rangle_{CAB} =   
|+\rangle|+\rangle|\psi^+(t)\rangle ~-~ 
|-\rangle|-\rangle|\psi^-(t)\rangle ,
\end{equation}
in place of Eq. (\ref{diftau}).
Now Bob first measures $C$ at time $t_1$ in the $+/-$ basis. 
Thereby he disentangles all three qubits and instantaneously
knows whether Alice's qubit is left in the state $|+\rangle$ or $|-\rangle$
without Alice's classical communication.

\begin{figure}
\centerline{\psfig{figure=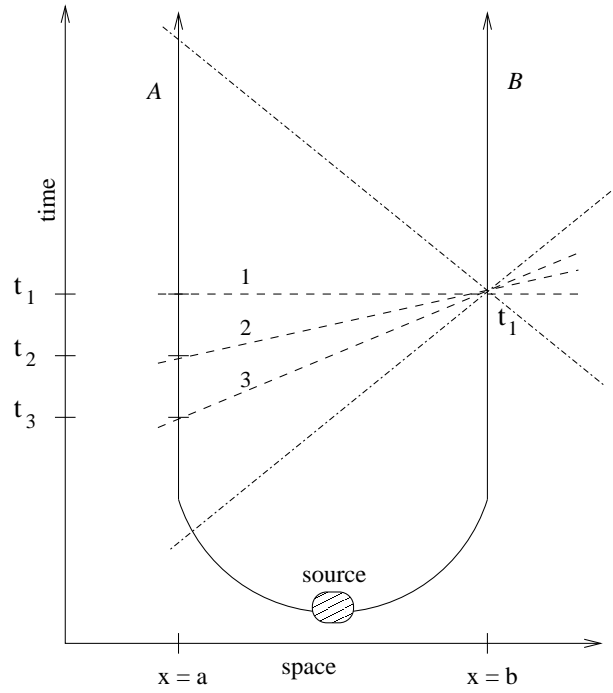,width=8.0cm}}
\caption{$A$ and $B$ are two entangled particles, located at $x = a, b$,
respectively, whose most probable
worldlines are indicated by the bold lines. The dashed
lines 1, 2 and 3 are possible spacelike collapse hypersurfaces corresponding 
to Bob's measurement on Alice's proxy qubit at event $(b, t_1)$, and
intercepting Alice's worldline at times $t_1$, $t_2$ and $t_3$. The choice of
the collapse hypersurface determines the phase of Alice's quantum clock at
the instant of disentanglement, and hence the correlations between qubits
$B$ and $C$ observed 
by Bob. The dash-dotted lines represent the future and past lightcones 
of event $(b, t_1)$.}
\label{korski}
\end{figure}

In traditional parlance, Bob's measurement on $C$
collapses (or ``decoheres"/``reduces") 
the $CAB$ system along his equal-time hypersurface. In Figure
\ref{korski}, this is the surface 1, which intercepts $A$'s worldline at
$t_a = t_1$, i.e., at event $(a, t_1)$. For each shared pair of qubits,
if he finds $C$ (and hence, $A$) in the state $|\pm\rangle$,
he concludes that $B$ is left in the state $|\psi^{\pm}\rangle$.
After performing a second measurement, on $B$, he expects to find the 
correlations in Eq. (\ref{probs})
with $t_a = t_1$. On the other hand, a third observer, whose 
equal-time hypersurface passing through event $(b, t_1)$ is given by surface
2 in Figure \ref{korski}, would expect that Alice's qubit disentangles at event
$(a, t_2)$ and that Bob would find correlations given by Eq. (\ref{probs}) 
with $t_a = t_2$, rather than $t_a = t_1$. Similarly, 
correlations with $t_a = t_3$ in Eq. (\ref{probs}) are predicted to be found 
by an observer whose equal-time hypersurface is
given by surface 3 in Figure \ref{korski}, and so on. This
implies that in order to ensure the Lorentz invariance of Eq. (\ref{probs}),
a unique hypersurface should correspond to Bob's measurement, 
which can inferred by Bob from the observed
correlations. Tests of this kind can be implemented at present, e.g.,
using rapidly moving detectors (cf. Refs. \cite{sca00zbi00}).

We wish to stress that the preceding result does not violate causality,
since no classical signal is transferred along the hypersurface. Only
correlations are affected. Indeed, in one sense the
dependence of correlations on the choice of hypersurface, and thence the need
for a covariant description of observables based on a singled out hypersurface,
brings quantum measurement closer 
to the spirit of Special Relativity. A more detailed discussion of the above
experiment and repercussions of our result for the epistemology and 
models of quantum measurement and for quantum information and relativistic
quantum mechanics will be dealt with in future works. 

{\small I thank Mr. B. Sahoo and Dr. D. Chakalov for discussions and
suggestions.}

\end{document}